\newif\ifaastex\aastexfalse

\ifaastex
\documentclass[12pt, preprint]{aastex}
\else
\documentclass{emulateapj}
\fi

\usepackage{natbib}
\ifaastex
\else
\let\jjdagger=\dagger
\let\jjddagger=\ddagger
\renewcommand{\dagger}{\ensuremath{\jjdagger}}
\renewcommand{\ddagger}{\ensuremath{\jjddagger}}
\fi

\newcommand{\theSNR}{SNR~0104-72.3}

\newcommand{\hi}{\ion{H}{1}\ }
\newcommand{\halpha}{H$\alpha$}

\newcommand{\hms}[3]{#1$^{\mathrm h}$ #2$^{\mathrm m}$ #3$^{\mathrm s}$}

\newcommand{\dms}[3]{#1$\arcdeg$ #2$\arcmin$ #3$\arcsec$}

\newcommand{\um}{\ensuremath{\mu\mathrm{m}}}

\newcommand{\MYNOTE}[1]{{{#1}}}

\shorttitle{Chandra study of SNR 0104}
\shortauthors{Lee et al.}

\begin{document}


\title{\theSNR: A remnant of Type Ia Supernova in a Star-forming region?}

\author{Jae-Joon Lee\altaffilmark{1,2} Sangwook Park\altaffilmark{3},
{John P. Hughes\altaffilmark{4}},
{Patrick O. Slane\altaffilmark{5}},
and {David N. Burrows\altaffilmark{6}}}

\altaffiltext{1}{Korea Astronomy and 
Space Science Institute, Daejeon 305-348, Korea}
\altaffiltext{2}{leejjoon@kasi.re.kr}
\altaffiltext{3}{Department of Physics, University of Texas at Arlington
Arlington, TX 76019}
\altaffiltext{4}{Department of Physics and Astronomy, Rutgers University, 136
Frelinghuysen Road, Piscataway, NJ 08854-8019}
\altaffiltext{5}{Harvard-Smithsonian Center for Astrophysics, 60 Garden Street,
Cambridge, MA 02138}
\altaffiltext{6}{Astronomy and Astrophysics Department, Pennsylvania
  State University, University Park, PA 16802}





\begin{abstract}
  We report our 110 ks Chandra observations of the supernova remnant
  (SNR) 0104-72.3 in the Small Magellanic Cloud (SMC). The X-ray
  morphology shows two prominent lobes along the northwest-southeast
  direction and a soft faint arc in the east.  Previous low resolution
  X-ray images attributed the unresolved emission from the
  southeastern lobe to a Be/X-ray star.  Our high resolution Chandra
  data clearly shows that this emission is diffuse, shock-heated
  plasma, with negligible X-ray emission from the Be star.
  The eastern arc is positionally coincident
  with a filament seen in optical and infrared observations. Its X-ray
  spectrum is well fit by plasma of normal SMC abundances, suggesting
  that it is from shocked ambient gas.  The X-ray spectra of the lobes
  show overabundant Fe, which is interpreted as emission from the
  reverse-shocked Fe-rich ejecta.  The overall spectral
  characteristics of the lobes and the arc are similar to those of
  Type Ia SNRs, and we propose that \theSNR\ is the first case for a
  robust candidate Type Ia SNR in the SMC. On the other hand, the
  remnant appears to be interacting with dense clouds toward the east
  and to be associated with a nearby star-forming region. These
  features are unusual for a standard Type Ia SNR. Our results 
  suggest an intriguing
  possibility that the progenitor of \theSNR\ might have been a white
  dwarf of a relatively young population.
\end{abstract}






\section{Introduction}

Among about a dozen supernova remnants (SNRs) detected in the Small
Magellanic Cloud (SMC), only three X-ray brightest SNRs (0102-72.3,
0103-72.6, and 0049-73.6) have been relatively well studied. Because
all these three SNRs most likely originate from a core-collapse SN,
revealing the type of other SNRs is of great importance to study the
star-forming history, SN rates, and chemical evolution of the
SMC. \theSNR\ is the fourth brightest X-ray SNR in the SMC, whose
origin has been controversial.
Its optical emission was suggested to be Balmer-dominated
\citep{1984ApJS...55..189M}, indicating Type Ia origin. On the other
hand, with the ROSAT data, \citet{1994AJ....107.1363H} identified an
unresolved X-ray source in the southern part of the SNR as a candidate
Be/X-ray star which appeared to be co-spatial with the SNR. 
Based on this plus other evidence for a Population I
environment, they suggested a core-collapse origin for
\theSNR.
With the XMM-Newton data, \citet{2004A&A...421.1031V} found evidence
for enhanced Fe L line emission from the remnant, with which they
reinstated Type Ia origin. Recently, with Akari IRC observations,
\citet{2007PASJ...59S.455K} discovered bright infrared shells which
are positionally coincident with the optical \halpha\ filaments
surrounding the X-ray emission \citep{1994AJ....107.1363H}. IR color
distributions indicated shock interaction with ambient molecular
clouds, which may support a core-collapse explosion of a massive
progenitor star for the origin of \theSNR.



We report here on the results from our observations of \theSNR\ with
the \emph{Chandra} X-Ray observatory. We note that the remnant has
been serendipitously detected a number of times by \emph{Chandra}
during the ACIS calibration observations of SNR 0102-7219. However,
\theSNR\ was detected mostly in FI CCDs at large off-axis angles.  Due
to the low sensitivity and the poor spatial resolution, these archival
data were not suitable for our analysis and we only report results
based on our new observations.

\section{Observation \& Results}
\label{sec:observation}

We observed \theSNR\ with the Advanced CCD Imaging Spectrometer
(ACIS) on board \emph{Chandra} on 2008 Jan 27--31 during AO9.  We used
ACIS-S3 for the effective detection of the soft X-ray emission of the
SNR. A total of 110 ks of exposure, with two separate obsIDs (54 ks
for obsID 9100, 56 ks for obsID 9810) was obtained in the VFAINT mode.
The level 1 event files were reprocessed to create new level 2 event
files. We applied parameters of the standard \emph{Chandra} pipeline process,
except that we applied the VFAINT mode background cleaning and turned
off the pixel randomization. CIAO 4.2 and CALDB 4.2.2 were
used for all the reprocessing and analysis. We examined the overall
background light curve for periods of high background.
No significant background flare was seen and we did not apply any light
curve filtering.

\begin{figure*}
  \centering
  \plotone{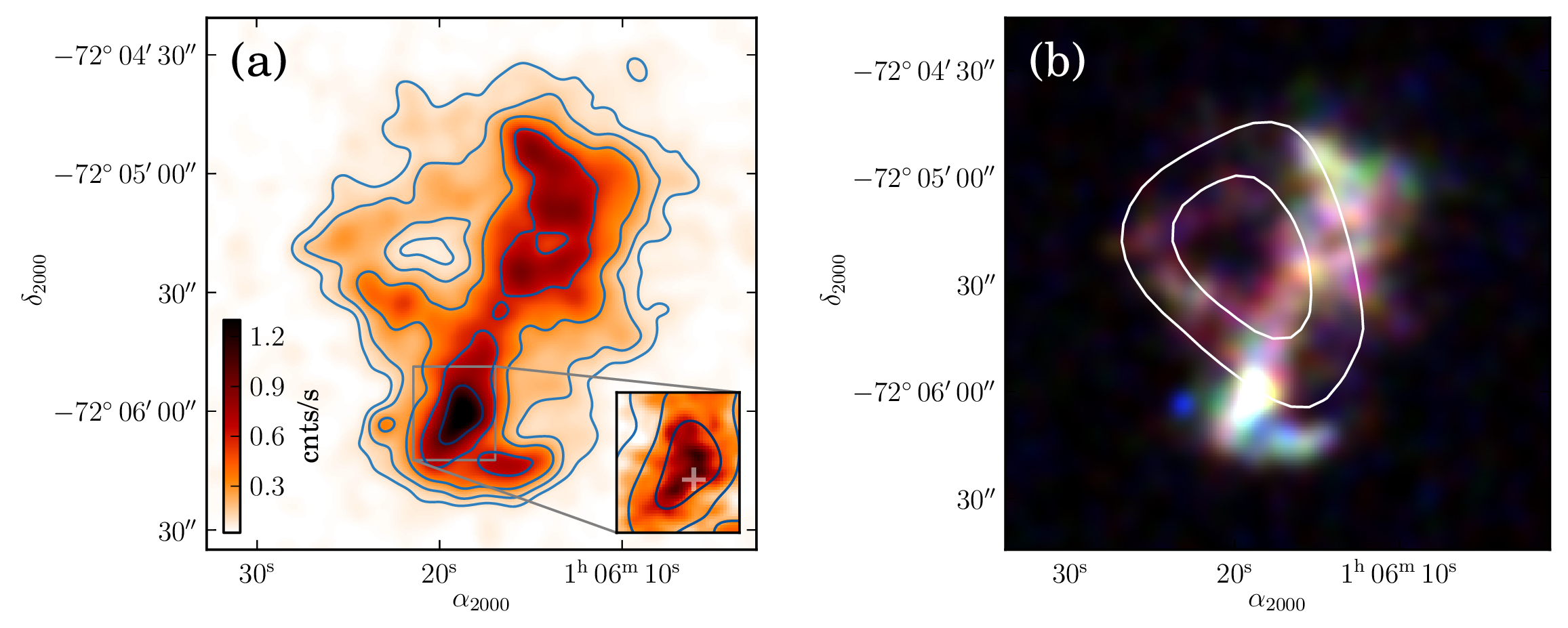}
  \caption{ (a) Chandra image (0.6-3.0 keV) of \theSNR\ smoothed with
    Gaussian beam of 2.5\arcsec. (inset) zoom-up of the bright
    southern lobe, where previous observations suggested as a point
    source, smoothed with 1\arcsec\ beam. The location of the candidate
    Be star is marked with a white plus sign. (b) RGB composite image
    of the Chandra data overlaid with white contours of the 843 MHz
    radio map \citep{1999AJ....117.1578B}. Red is 0.6-0.95 keV,
    green is 0.95-1.12 keV, and blue is 1.12-3 keV bands.}
  \label{fig:chand}
%
%
  \label{fig:rgb}
\end{figure*}

\section{X-ray Images}
\label{sec:imaging-anaylsis}

\begin{figure*}
  \centering
  \plotone{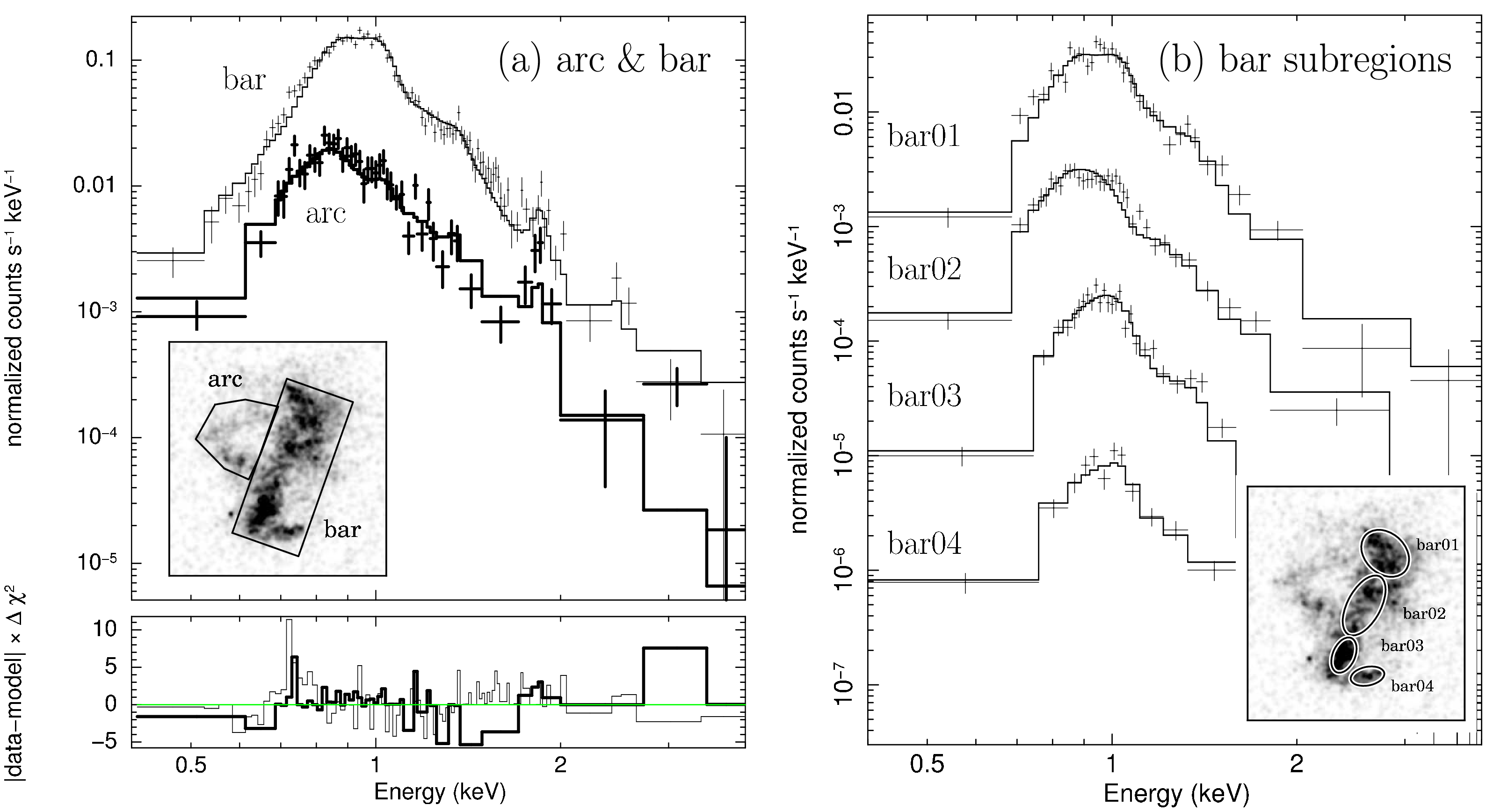}
  \caption{(a) Chandra spectra of the bar and the arc regions.  The
    inset image shows the extraction regions of the spectra. The solid
    lines are best fit models in Table~\ref{tab:fit}. The spectra are
    binned for display purpose only. (b) Chandra spectra of subregions
    of the bar. The inset image shows the extraction regions of the
    spectra. The spectra of bar02, bar03, bar04 are scaled by
    $10^{-1}$, $10^{-2}$ and $10^{-3}$, respectively for a display
    purpose. The solid lines are best fit models in
    Table~\ref{tab:fit}, scaled accordingly.}
  \label{fig:bars-comp}
\end{figure*}

Figure~\ref{fig:chand}(a) is the broad band (0.6-3.0 keV) Chandra ACIS
image of \theSNR. Our deep ACIS image reveals details of the X-ray
structures that have not been available from the previous ROSAT
and XMM-Newton observations
\citep{1994AJ....107.1363H,2004A&A...421.1031V}.  The X-ray morphology
of the remnant is dominated by bright lobes extending about 90\arcsec\
along the northwest-southeast direction.  Faint emission surrounding
the lobes is also clearly seen. Most notably, there is a faint
arc-like structure on the eastern side of the lobes.  We refer to the region
including two lobes as the ``bar'' region, and the faint arc emission
in the east as the ``arc'' region (see inset image of
Figure~\ref{fig:bars-comp}(a)).  Figure~\ref{fig:rgb}(b) shows a RGB
composite image of \theSNR.
The bar and the arc regions show different colors in the RGB
image. The arc is dominated by red while the bar shows a complex
mixture of three colors.
\MYNOTE{
A color variation is also noticed within the bar, where the bright
southern and northern regions of the bar are more greenish
(i.e., 0.95-1.12 keV) and/or blueish (i.e., 1.12-3 keV) than the
regions in the middle.  There is thus a trend of higher photon
energies toward the ends of the bar than in the middle.
}

\section{Spectral Analysis}
\label{sec:spectral-analysis}

The high-resolution ACIS imaging observations allow us to perform a
spatially resolved spectral analysis of \theSNR. We extracted X-ray
spectra from several characteristic regions, based on the color
variation in the RGB composite image.  For a background subtraction,
we used an average spectrum from four circular regions ($\sim
30\arcsec$ in radius) around the SNR.  For spectral modeling, 
unbinned spectra were fit using the Churazov weighting
\citep{1996ApJ...471..673C}.

We first extract spectra from
the arc and the bar. The background-subtracted spectra are shown in
Figure~\ref{fig:bars-comp}.
The observed spectra are soft, with most of the emission seen below 3
keV. The peak energy of the arc is lower than that of the
bar, which explains the different color between the arc and the
bar in Figure~\ref{fig:rgb}(b).
%
We fit the spectra with a single nonequilibrium ionization (NEI)
plane-parallel shock model \citep[][vpshock in Xspec
v12]{2001ApJ...548..820B} that is based on ATOMDB
\citep{2001ApJ...556L..91S}.  We use an augmented version of this
atomic database to include inner-shell processes and updates of the Fe
L-shell lines.\footnote{The augmented APEC atomic data have been
  provided by K. Borkowski. The relevant discussion on these detailed
  plasma model issues has been presented in
  \citet{2006ApJ...645.1373B}.}  
\MYNOTE{
We modeled the foreground absorption
by the Galaxy and the SMC separately.  For the absorption by the
Galaxy, We used a fixed column density of $N_{\mathrm{H,Gal}} = 2.2
\times 10^{20}$ cm$^{-2}$ which is the \hi\ column density of the
Galaxy toward \theSNR\ \citep{1990ARA&A..28..215D}.  The foreground
column density in the SMC ($N_{\mathrm{H,SMC}}$) is fit as an additional
absorption component assuming the SMC interstellar abundances
\citep{1992ApJ...384..508R}.  
}
We first try to fit the spectra with
elemental abundances 
of the emitting gas fixed at those of the SMC values
\citep{1992ApJ...384..508R}. 
An acceptable fit is obtained for the spectrum of the arc
region. 
The best fit results give hydrogen column density of
$N_{\mathrm{H,SMC}} \sim 1 \times 10^{22}$ cm$^{-2}$ and an electron
temperature of 0.54 keV, although the ionization time scale is poorly
constrained. This is the region where the remnant is thought to
interact with dense ambient gas (see \S~\ref{sec:discussion}), and the
X-ray spectral fit result is consistent with such an interpretation,
i.e., the X-ray emission of the arc is predominantly from shocked gas
with SMC abundances.  On the other hand, the spectrum of the bar
region cannot be adequately fit with models using SMC abundances
(reduced chi-squared $\sim3$); the prominent peak near 1 keV, in
particular, could not be reproduced.  The strong lines in this energy
range are those of Ne K and Fe L shell lines.
%
Freeing abundances of Ne and Fe provides an acceptable fit.  The best
fit model has overabundant Fe (by a factor of 10 increase above the SMC
value) and negligible amount of Ne.  We note that quantitative
interpretations of the Fe abundance are difficult due to several
modeling issues including incomplete Fe L line modeling
\citep[e.g.,][]{2006ApJ...645.1373B}, and we interpret this number
only in general sense of the Fe-rich ejecta.
Our results of the Fe overabundance is consistent with the results of
\citet{2004A&A...421.1031V}.  
\MYNOTE{
The \hi\ column density of the SMC
toward \theSNR\ is $\sim 6\times 10^{21}$ cm$^{-2}$
\citep{1999MNRAS.302..417S}.  While the estimated $N_{\mathrm{H,SMC}}
\sim 9 \times 10^{21}$ cm$^{-2}$ is slightly larger than the \hi\ column
density, we note that HI column density values typically underestimate the
total hydrogen column density.
}


The RGB composite image in Fig.~\ref{fig:rgb}(b) shows varying colors
within the bar region, indicating spectral variations although some of
them may be simply due to low photon statistics.
Figure~\ref{fig:bars-comp}(b) shows extracted spectra from 4 smaller
regions inside the bar, and spectra of some regions (e.g., bar04) show
slightly different spectra than others. We fit spectra from each
subregion with the vpshock model.  Because of the poor photon
statistics of individual regions, we could not constrain model
parameters with both Ne and Fe abundances varied in the fits.
Thus, as the previous fit results of the bar suggested negligible
contribution of Ne, we fix Ne at the SMC abundance (fixing it at the
best fit parameter of the bar region has only minor effect on the
results).  We further fix the absorbing column density at the best fit
parameter of the bar region ($N_{\mathrm{H,SMC}} = 9.1 \times 10^{21}
\mathrm{cm}^{-2}$). This approach provides an acceptable fit with
reasonable parameter constraints.  We find that the spectra from the
central regions tend to have lower temperature and lower Fe abundance.
The origin of this spectral variation is unclear with the current
data. We speculate that the thermal conditions and/or metal abundances
may be inhomogeneous along the bar region, possibly due to the mixture
of the ejecta and ambient medium and/or intrinsically non-uniform
distribution of the ejecta.  And the spectral fits suggest that the
central regions may be less dominated by ejecta material.  All the fit
results are tabulated in Table~\ref{tab:fit}.

\ifaastex
\begin{deluxetable}{cccccccc}
\else
\begin{deluxetable*}{cccccccc}
\fi
\tablecolumns{8}
\tablewidth{0pc}
\tablecaption{Fit Results\label{tab:fit}}
\tablehead{
\colhead{Region} & \colhead{$N_\mathrm{H,SMC}$} &
\colhead{kT} & \colhead{Ne} & \colhead{Fe} & \colhead{$\log \tau$} &
\colhead{norm} & \colhead{reduced $\chi^2$}\\
\colhead{} & \colhead{[$10^{21}$ cm$^{-2}$]} &
\colhead{[keV]} & \colhead{} & \colhead{} & \colhead{} &
\colhead{[$10^{-4}$]} & \colhead{}
}
\startdata
{Arc} & {9.9$_{-2.9}^{+0.22}$} & { 0.54$_{-0.05}^{+ 0.05}$} 
      & {0.20} & {0.13} 
      & {$> 12.5$} & { 1.4$_{-0.7}^{+0.3}$ } & {267/243}\\
{Bar} & {9.1$_{-1.7}^{+1.1}$} & { 1.95$_{-0.09}^{+ 0.15}$} 
      & {$<0.04$} & { 1.26$_{-0.15}^{+ 0.11}$} 
      & { 11.0$_{-0.1}^{+0.1}$} & { 0.85$_{-0.09}^{+0.05}$} & {275/241}\\
{Bar01} & {$9.1$} & { 2.04$_{-0.24}^{+ 0.41}$} 
      & {0.20} & { 1.65$_{-0.35}^{+ 0.64}$} 
      & { 11.1$_{-0.1}^{+0.1}$} & { 0.14$_{-0.04}^{+0.03}$} & {280/243}\\
{Bar02} & {$9.1$} & { 0.72$_{-0.03}^{+ 0.04}$} 
      & {0.20} & { 0.51$_{-0.13}^{+ 0.20}$} 
      & { $> 12$ } & { 0.36$_{-0.08}^{+0.09}$} & {206/243}\\
{Bar03} & {$9.1$} & { 1.41$_{-0.29}^{+ 0.25}$} 
      & {0.20} & { 3.4$_{-0.8}^{+ 3.0}$} 
      & { 11.1$_{-0.1}^{+0.2}$} & { 0.07$_{-0.03}^{+0.03}$} & {199/243}\\
{Bar04} & {$9.1$} & { 1.78$_{-0.60}^{+ 2.30}$} 
      & {0.20} & { 0.68$_{-0.19}^{+ 0.34}$} 
      & { 11.2$_{-0.1}^{+0.1}$} & { 0.09$_{-0.03}^{+0.17}$} & {213/243}\\
\enddata
\tablecomments{The errors are shown in 90\% confidence range. Unless
  indicated with uncertainty values derived from the fits, the metal
  abundances are fixed at those of the SMC values from
  \citet{1992ApJ...384..508R}. They are He = 0.05, C = 0.13, N = 0.20,
  O = 0.10, Ne = 0.20, Mg = 0.12, Si = 0.18, S = 0.15, Ar = 0.08, Ca =
  0.20, Fe = 0.13, Ni = 0.20. All values are relative to the solar
  ones. The $N_\mathrm{H,SMC}$ values for the individual bar regions
  (Bar01, Bar02, etc) are
  fixed at the value obtained for the entire bar.} 
\ifaastex
\end{deluxetable}
\else
\end{deluxetable*}
\fi

\begin{figure*}
  \centering
  \plotone{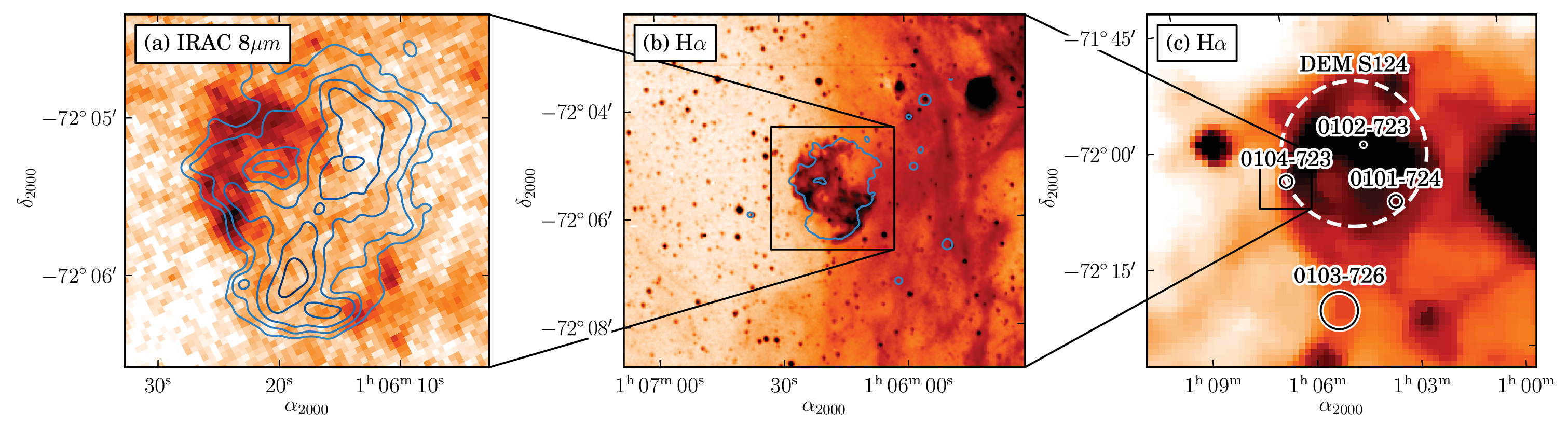}
  \caption{(a) Infrared image of \theSNR\ (IRAC 8\um\ of Spitzer
    archival data) with the contours of Chandra broadband image in
    Figure~\ref{fig:chand}(a). (b) \halpha\ image from
    \citet{1994AJ....107.1363H}. The lowest level contour line from (a)
    is shown. (c) \halpha\ image around \theSNR\ from
    the Southern \halpha\ Sky Survey Atlas
    \citep{2001PASP..113.1326G}. The solid circles are locations of
    known SNRs. The dashed circle is the approximate extent of the
    supperbubble DEM~S124. }
  \label{fig:ir-comp}
\end{figure*}

\section{Discussion}
\label{sec:discussion}

\subsection{The SN type of \theSNR}

Based on the ROSAT HRI observation, \citet{1994AJ....107.1363H}
attributed the unresolved X-ray emission feature at $\alpha_{2000}$,
$\delta_{2000}$ = \hms{01}{06}{18}, \dms{-72}{06}{00} to emission
associated with a candidate Be star, interpreting it as a Be/X-ray
star.
%
%
The arcsecond resolution and good photon statistics of our deep
Chandra data reveal that this X-ray emission feature is extended
(Figure~\ref{fig:chand}), and no point-like source is found within a
few arcsecond from the position of the Be star candidate (on-axis
astrometric uncertainties of the ACIS is $\sim 0.6\arcsec$). We
examined the RGB composite image and various narrow band images of
different energy ranges and their ratio images, but could not find any
point-like feature that is positionally associated with the Be
candidate.  The overall X-ray spectrum of this region is also similar
to X-ray spectra in other regions. Thus we conclude that the X-ray
emission in this region originates from the shocked SNR gas, and not
from the point source.  
Although the Be star that \citet{1994AJ....107.1363H} identified
is unlikely to be a significant X-ray emitter, the possibility
that \theSNR\ and the Be star belong to the same
OB association, as they also suggested, remains valid,
but would then argue against a typical Type Ia remnant. This
will be further discussed  in \S~\ref{sec:env}.


\label{sec:comp-ir}

The overabundance of Fe in the bar region suggests that \theSNR\ is
likely the remnant of Type Ia SN.  Indeed, the spectra of the bar
regions are quite similar to ejecta spectra found in the well-known
Type Ia SNR DEM~L71 \citep{2003ApJ...582L..95H} in the Large
Magellanic Cloud (LMC) and several other relatively old Type Ia LMC
SNRs \citep[e.g.,][and references therein]{2006ApJ...652.1259B}.
DEM~L71's central X-ray emission, which is well described by shocked
Fe-rich ejecta, is similar to that of \theSNR.
DEM~L71 also shows faint soft emission of normal LMC abundances
surrounding the central emission, similar to the arc in \theSNR.
%
Therefore, based on the overabundant Fe from our spectral fit and the
spectral similarity of the bar and the arc region to other Type Ia
SNRs, we propose that \theSNR\ is the first solid candidate for Type Ia
SNR in the SMC.
\MYNOTE{
Another element expected to be a signature of Type Ia SNR is Si 
\citep[e.g.,][]{2007ApJ...662..472B}. 
Allowing the Si abundance to vary in the fit yields a slightly
larger value, but still comparable to the SMC abundance within the fit
uncertainty.
}
The Type Ia origin is supported by non-detection of a pulsar and/or
pulsar wind nebula, although they could be too faint to be detected as
in other core-collapse SNRs in the SMC
\citep[e.g.,][]{2003ApJ...598L..95P}.


\subsection{Type Ia SNR in Unusual Environment?}
\label{sec:env}

Our deep Chandra observations find that the X-ray emission from the
bar region is prominently from Fe-enriched SN ejecta of a Type Ia SNR.
%
Based on limited sample, the X-ray morphologies of young Type Ia SNRs
have been claimed to be statistically more symmetric than
core-collapse SNRs \citep{2009ApJ...706L.106L}.  
Although a quantitative
analysis of the asymmetry in the morphology of \theSNR\ should be
performed (which is beyond the scope of this work), the highly
elongated morphology of \theSNR\ appears to be inconsistent with
the results by \citet{2009ApJ...706L.106L}.
Such an elongated Fe-rich ejecta feature is similar to that found in
the Galactic SNR W49B.  Although the type is uncertain for W49B
\citep[e.g.,][]{2006A&A...453..567M}, the highly elongated morphology
of ejecta may suggest
an asymmetric explosion. Alternatively, given the possibility that the
remnant is interacting with dense material toward the east (see
below), the atypical morphology could be due to highly inhomogeneous
ambient environment of the remnant.

The environment around \theSNR\ is also unusual for a Type Ia SNR.
The relatively high absorbing column density toward \theSNR\ indicates
that the remnant could be located at dense environment.  
The eastern X-ray arc is positionally
coincident with
an optical and infrared shell (Figure~\ref{fig:ir-comp}).  
The optical emission shows dominant
\halpha\ line with weak lines of other ions such as [\ion{O}{3}] and
[\ion{S}{2}], and \citet{2007MNRAS.376.1793P} found that their line
ratios are consistent with radiative shocks in other SNRs.  
\citet{2007PASJ...59S.455K} reported infrared emission associated with
\theSNR\ from their \emph{Akari} observations.  The observed infrared
colors were consistent with those from shocked molecular clouds, and
\citet{2007PASJ...59S.455K} suggested that \theSNR\ is probably
interacting with molecular clouds.  In this regard, the radio
continuum emission of \theSNR\ could be mostly from the shocked
ambient gas.  
Although the radio
data quality is poor, Figure~\ref{fig:chand}(b) shows that the extent of the radio
continuum covers
the
eastern arc region, and also the central region of the bar where the
Fe overabundance appears to be less prominent.
Thus, 
observations seem to provide a consistent picture that the arc is
where the remnant is interacting with dense ambient (possibly
molecular) clouds.  On the other hand, the NW and SE parts of the SNR,
where only the X-ray emitting ejecta are visible and there is no
optical or IR emission, probably corresponds to low density regions
where the forward shock in the ISM is still relatively fast and
radiative shocks have not formed.

In fact, the projected location of \theSNR\ is at the eastern boundary
of the superbubble DEM S124 \citep{1976MmRAS..81...89D} (see
Figure~\ref{fig:ir-comp}).  
In Figure~\ref{fig:ir-comp}(c),
we show the \halpha\ emission complex toward the direction of \theSNR\
and overlay locations of known SNRs in the SMC. It appears that
\theSNR\ and SNR~0102-7219, a well-known core-collapse SNR, belong to
the same star-forming region.
The Type~Ia origin inferred from the Fe-rich ejecta and the
association of the remnant with the star-forming region are not
consistent with conventional scenarios of standard Type Ia, which
involve the
explosion of an old C/O white dwarf in a binary.
%
Recent studies of a large sample of Type Ia SNe suggested that
progenitors of some Type Ia SNe (``prompt'' type) could have been
relatively young massive stars with high metallicity
\citep[e.g.,][]{2005ApJ...629L..85S,2008A&A...492..631A}.
\MYNOTE{
More recently, from the systematic study of the SNRs in the Magellanic
Clouds, \citet{2010MNRAS.407.1314M} suggested that some of the SNRs
could be remnants of the prompt Type Ia SNe.
}
While we cannot completely rule out the possibility of mere
coincidence, the unusual nature of \theSNR\ (i.e., Type Ia
characteristics of Fe-rich X-ray ejecta in the SNR and the SNR's
spatial association with an environment with a strong star-forming
activity) suggests an intriguing possibility that \theSNR\ may be a
candidate SNR of a prompt SN Ia.

\acknowledgements
This work was supported in part by the Smithsonian Astrophysical Observatory
under Chandra grant GO8-9509A. P.O.S. acknowledges partial support from NASA
contract NAS8-03060.


\end{document}